*Research Article*

# Application of Blockchain and Internet of Things to Ensure Tamper-Proof Data Availability for Food Safety

**Adnan Iftekhar** ,[1] **Xiaohui Cui** ,[1] **Mir Hassan**,[2,3] and **Wasif Afzal**[4]

[1]Key Laboratory of Aerospace Information Security and Trusted Computing, Ministry of Education,
 School of Cyber Science and Engineering, Wuhan University, Wuhan, China
[2]Wuhan National Laboratory for Optoelectronics, Huazhong University of Science and Technology, Wuhan, China
[3]University of Glasgow, Glasgow, UK
[4]Mälardalen University, Västerås, Sweden

Correspondence should be addressed to Xiaohui Cui; xcui@whu.edu.cn

Academic Editor: Luis Patarata

Food supply chain plays a vital role in human health and food prices. Food supply chain inefficiencies in terms of unfair competition and lack of regulations directly affect the quality of human life and increase food safety risks. This work merges Hyperledger Fabric, an enterprise-ready blockchain platform with existing conventional infrastructure, to trace a food package from farm to fork using an identity unique for each food package while keeping it uncomplicated. It keeps the records of business transactions that are secured and accessible to stakeholders according to the agreed set of policies and rules without involving any centralized authority. This paper focuses on exploring and building an uncomplicated, low-cost solution to quickly link the existing food industry at different geographical locations in a chain to track and trace the food in the market.

## 1. Introduction

Food supply chains are vast and spread over the whole world. These supply chains connect three significant sectors of an economy. The agriculture sector includes crops and livestock, the food-manufacturing industry, and the marketing sector, which provides distribution, wholesales, and retails. The three significant sources of food are crops, livestock, and seafood. We are contaminating our crops with pesticides and excess fertilizers. The use of growth hormones and the administration of drugs is becoming normal to increase milk and meat production. It is severely affecting human health and increases the risk of various cancers in humans [1]. The physical contaminants with foreign materials, persistent organic pollutants, tampering BB (best before) dates, altering documentation, misprinting ingredients, and inappropriate storage temperatures are some significant sources raising food safety and public health risks during processing and shipment process [2].

Health consciousness is also an increasing phenomenon in recent years. As consumers are becoming more and more health-conscious, the demand for certain production methods to meet specific environmental, ethical, and nutritional needs is increasing [3]. Meat and milk producers are introducing organic, natural, and grass-fed production in the market [4]. These livestock farmers and producers are claiming their meat and milk production more nutritional and superior than traditionally produced meat and milk. In the organic meat and milk production system, the animals are raised on 100% certified organic food with strictly controlled vaccination. The basic similarities and differences between traditional, natural, grass-fed, and organic production systems are summarized in Table 1. These organic and safety certified products are becoming more prominent in markets as a trend towards purchasing organic food is growing among consumers [5]. This emergence of new markets and a global increase in food prices also increase the growth of food frauds and food safety risks [6].

Table 1: Similarities and differences between different meat and milk production systems.

| | Traditional harvest system | Natural harvest system | Grass-fed harvest system | Organic harvest system |
|---|---|---|---|---|
| Food | Pasture in grazing system, high-concentrate diet | Minimally processed, no artificial ingredients, no preservatives | Fed on grass and grains, none forage supplements | Certified organic, animals are raised in healthier facilities |
| Health management | Vaccinations, antibiotics, ionophores, growth hormones may be used | Vaccinations, limited antibiotics, no ionophores, no growth hormones | Vaccinations, no antibiotics used, no ionophores, no growth hormones | Only strictly controlled primary vaccination allowed |
| Marketing | Mainly auction markets | Some level of verification is required | Source and management verification are required as packer, wholesaler, retailer | Complete verification required from birth to retailer |

Food fraud refers to a group of activities that perform intentionally or unintentionally for economic gain. Spink and Moyer [2] defined food fraud as "Food fraud is a collective term used to encompass the deliberate and intentional substitution, addition, tampering, or misrepresentation of food, food ingredients, or food packaging, or false or misleading statements made about a product, for economic gain." The authenticity of the food and food labeling is becoming a significant concern for manufacturers, regulatory authorities, and consumers [7]. The food manufacturers and distributors are tampering to substitute or alter the product ingredients with inferior ones in order to set an appropriate price for the targeted market or yield more significant profits. This phenomenon is proving a fatal threat to human health. In 2008, a Chinese milk scandal found infant milk formula contaminated with melamine, which affected 0.3 million infants, of which about 54,000 were hospitalized, and six died due to kidney stones and other related problems [8]. In 2013, the beef burgers in Britain and Ireland were found to contain horse meat [9]. In 2014, a Chinese crime syndicate was arrested who sold about more than 1 million USD meat of rats, minks, and other small mammals under cover of mutton [10]. Some food safety-related incidents in South Korea from 1998 to 2016 are also documented [11]. The recent pandemic of coronavirus disease (COVID-19) started in December 2019, believed to originate from an unregulated wild animal's meat market in Wuhan. At the time of writing this paper, it has infected 2 million peoples worldwide, which resulted in about 0.1 million deaths and billions of dollar economic loss. The Chinese Journal of Food Hygiene published a study in 2011, claiming that more than 94 million people fall ill due to foodborne illnesses, resulting in about 85,000 deaths each year [12]. The UK Government published a very detailed review and recommendation report on the integrity and assurance of food supply chain networks to protect United Kingdom consumers in July 2014. The report presented a national food crime control framework to protect consumers from food fraud and ensure food safety. This report discussed all the factors of food supply chains and highlighted the potential factors that cause food supply chain failure and fraud possibilities [13].

When it comes to supply chain management in terms of food safety, the visibility of the supply chain is an important issue. The food supply chains are more complex than the other supply chains. It is a big challenge to make sure the presence of associated data in the food supply chains from origin to the destination. These data are essential to prevent foodborne illness risks, food integrity issues, and issuing various food certificates. Aung and Chang [14] described the importance of traceability in the food supply chain concerning food safety and quality improvement.

Moreover, consumers are now more concerned about evidence that the products they are buying are produced in proper environmental facilities with acceptable specifications. We also observed a high demand for certified food and meat in supermarkets of Wuhan during the Covid-19 pandemic. It is increasing the need for producers to share essential information with food certifiers, whole-sellers, retailers, and consumers. NGOs and government regulatory authorities are also demanding more transparency, visibility, and traceability throughout the supply chain from source to retail. United Nation Global Compact defines traceability as "the ability to identify and trace the history, distribution, location, and application of products, parts, and materials, to ensure the reliability of sustainability claims, in the areas of human rights, labor (including health and safety), the environment, and anticorruption" [15]. To fulfill the demands of the consumers, the food industry is also getting more concerned about tracking their suppliers and supply chain. Researchers from Wuhan University and Huazhong Agriculture University have formed an alliance with various food-manufacturing companies, agriculture and livestock farm owners, and information technology providers in Wuhan, Hubei, to develop a system to trace the food from farms to the fork [16]. We will refer to this alliance as a consortium in this article.

The core purpose of this work is to merge the traditional supply chain management practices with the blockchain to trace a food package from farm to fork with unique identity for each food package while keeping it uncomplicated for the workers. It will keep the business transactions tamper-proof and accessible to stakeholders according to the business policies and agreed contracts of data sharing between the companies without involving any centralized authority for monitoring.

The rest of the article is organized as follows: Section 2 discussed the background of the blockchain and Internet of Things (IoT) technology. Section 3 presents the existing literature on the application of IoT and blockchain technology in the food supply chains. Section 4 presents the methodology we used for this work. Section 5 introduces the details of our traceability system design and demonstrates the architecture and functions of it using a real-world use case from the food industry in Hubei, China. We analyzed hardware cost in Section 6 and challenges and obstacles in Section 7. We make a brief conclusion of this work in Section 8. Finally, the future directions are given in Section 9.

## 2. Background

### 2.1. Blockchain.
Bitcoin originated from the world's first blockchain-based application [17]. It is a list of transactions that are accessible by many participants and secured using digital signatures and cryptography hash functions. This list is distributed across many systems over peer-to-peer networks in almost real time, which makes it practically impossible to change in previous transactions or makes it very easy to detect any illegal change in records [18]. The blockchain transactions store into sequentially ordered blocks. Each block holds several transactions, and a hash signature that links this block to the previous block acts like a pointer that forms a cryptographically secured chain of blocks called the blockchain, illustrated in Figure 1. The miners of the network repeatedly perform a function to solve a complicated mathematical puzzle to find a unique hash signature for each block which is a mathematical proof that the block is mined [19].

The main benefit of the blockchain is its quality of immutability, which makes it secure and as well as easy to audit trial [20]. The blockchain can be programmed to record virtually anything that is expressible in code. The enterprises are already adopting this technology, and others are moving towards this technology. In the manufacturing business, the supply chain is the most critical factor. In a typical supply chain scenario, multiple independent parties take part in moving payload from point A to point B, and they must track it to all destinations. The grain supply chain usually passes through multiple storages at multiple destinations shipped by multiple logistics from farmer to the end consumers. Figure 2 illustrates the stages of the grain supply chain from farms to retailers. A tamper-proof distributed ledger can record the travel of a specific batch of production that where, when, and who shipped or stored it or if it needs to be shipped somewhere in a specific time.

The blockchain is classified into the three categories as public, private, and permissioned [21]. Bitcoin is a classic example of a public blockchain where all the participants can join, read, and write data without any permission from any authority. Any participant can be part of the consensus process called mining [22]. The private blockchain is limited within an organization where the participants are known and trusted. The permissioned blockchain is an example of a group of companies or consortiums where participants are bound to some legal contract to get permission to access, read, and write the blockchain. The consensus process of permissioned blockchains is based on pre-elected nodes within the consortium. A summary of this classification is summarized in Table 2.

There are many blockchain development platforms to utilize blockchains in making secure and transparent transactions between the organizations [23]. A comparison

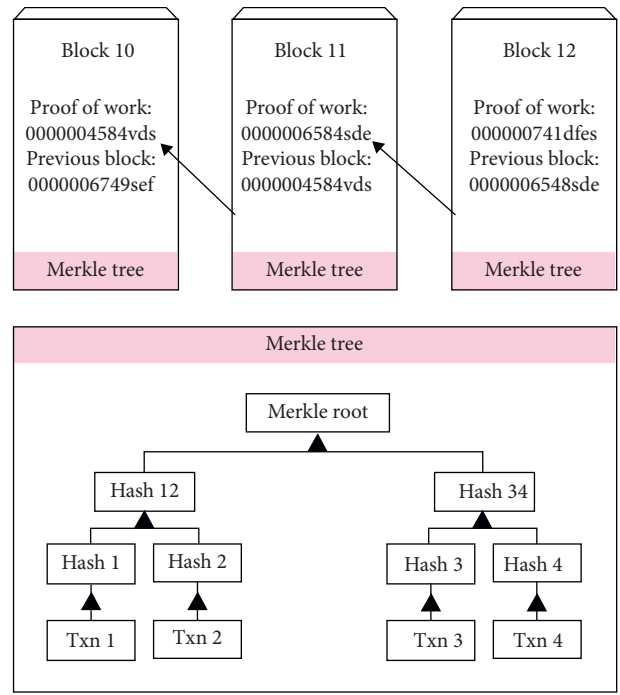

Figure 1: Blocks forming the blockchain using hash signature.

of some famous blockchain development platforms is given in Table 3.

### 2.2. Internet of Things.
Digitization is continuously growing and disrupting many aspects of our everyday life. The IoT platform enables the industry to have real-time tracking of assets and the environment. It also puts a critical impact on asset velocity, i.e., the assets linked to the inventory in the business world. The adoption of blockchain technology in the IoT industry is at its rise due to its provenance in security and tracking [24]. It is becoming essential to keep an eye on the animal's health and maintain batches with high efficiencies to increase the quality of the product and lower operating income. The demand for livestock identification and traceability increased the need for quality control and control of infectious diseases, medication, and its effects on the environment and consumer health [25, 26]. The recent advances and increasing phenomena of using radio-frequency identification (RFID) in society increased the standardization of RFID tag technology for specific purposes [27]. This phenomenon also expands the use of RFID-enabled biocapsules in the global livestock market. These capsules allow the farmer to capture the real-time data about body temperature, daily drinking cycles, ruminal pH level, and amount of activity among a large batch of cattle. These data can be gathered and monitored in real time with beacons across the farm. The farmer can receive the information anywhere through the web or mobile application. These data enable farmers to calculate the time and optimum insemination period after the estrus, which also prevents cattle from calving accidents. Furthermore, veterinary health

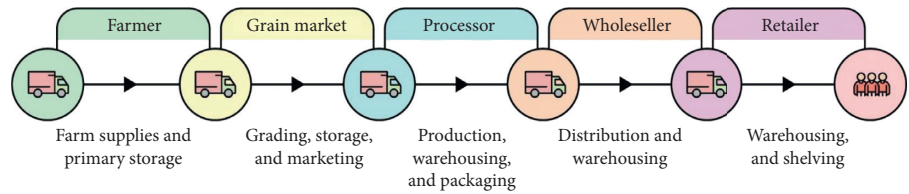

Figure 2: Storage and logistics are the backbone of an efficient supply chain.

Table 2: Classification of blockchains.

|  | Public blockchain | Private blockchain | Permissioned blockchain |
| --- | --- | --- | --- |
| Read access | No permission required from any authority | Read access is private within organization participants | Public/participants are permissible under some legal contracts |
| Write access | No permission required from any authority | Write access is private within organization participants | Participants are permissible under some legal contracts |
| Consensus process | Anyone can join consensus process | Pre-selected nodes within organization | Pre-selected nodes within consortiums |

Table 3: Comparison of blockchain development platforms.

|  | Fabric | VeChain | Ripple | R3Corda | Ethereum |
| --- | --- | --- | --- | --- | --- |
| Blockchain type | Permissioned | Public | Permissioned | Permissioned | Public and private |
| Cryptocurrency | None | VeThor Token | Ripple | None | Ether |
| Smart contract | Yes | Yes | No | Yes | Yes |
| Security model | Membership services | Block security protocol | Node validated and confirmed transactions | Permissioned only | Data are public and not encrypted |
| Integration efforts | Vary | Easy | Easy | Vary | Difficult |

services and owners are notified in real time with an alarm and message in case of any abnormality.

The 2013 horse meat scandal found 100% horse meat content in beef products in some cases [28]. The livestock farm animals were RFID identified. The beef originated from Doly-Com Romania was identified as horse meat [29]. This horse meat was delivered to a cold storage company in Breda by Draap Trading Ltd. Draap sold this meat to another company in Europe named Spanghero. Spanghero sold this beef like meat and insisted that it received the meat labeled as "'Beef." According to investigations reported by French media, Spanghero tampered the documents regarding the meat [30]. This was the failure of the traditional product and data flow system as illustrated in Figure 3.

## 3. Related Work

The blockchain is usable in medicine, economics, energy, and resource management. It is usable for exchanging almost everything that has digital representation. Dubai is becoming the first nation to use blockchains and the Internet to ensure food safety and consumer nutritional needs and priorities. Dubai government is about to digitize all the food items from farm to fork using blockchains and Internet of Things technology before Dubai expo 2020 [31]. In 2016, Walmart collaborated with IBM and Tsinghua University in order to launch a food safety collaboration center in Beijing to improve the tracking of food items in the supply chain with the help of blockchain technology. In 2017, Jingdong (jd.com) also joined this collaboration [32]. Jindong has also made a venture with an inner Mongolia-based beef producer, Kerchin, to apply blockchain technology in the production process. It enables its customers to track the information about the frozen meat, such as cow's breed, weight, and diet as well as the location of farms by just scanning the QR code available on the food package. It is also about to put more than 20 food items on the blockchain with Kerching as a supply chain partner [33]. ZhongAn Technology, a China-based technology company, developed a platform to track and record the chicken farming. They are also developing shared ledger-based business technologies and strategic solutions [34]. Alibaba collaborates with Blackmores and many other Australia and New Zealand-based food producers and suppliers to create a blockchain-based platform that combats the rise of counterfeiters targeting Australian and New Zealand-based food items sold across China on its platform [35]. IBM has teamed up with Krogen, McCormick and Company, McLane Company, Discall's Tyson Foods, Golden State Foods, Unilever, Nestle and Dole, and many others to implement distributor ledger technology [36].

Alzahrani and Bulusu have proposed a block-supply chain based on blockchain and near-field communication technology to tackle counterfeit products. The authors make sure that only the node that has the product will offer a new block, and the other nodes validate it. Second, the authors proposed a method to choose validators and a new consensus protocol based on Tendernint with the ability to select a different set of validators for each validation event. They simulated that their new protocol is very efficient for large networks [37].

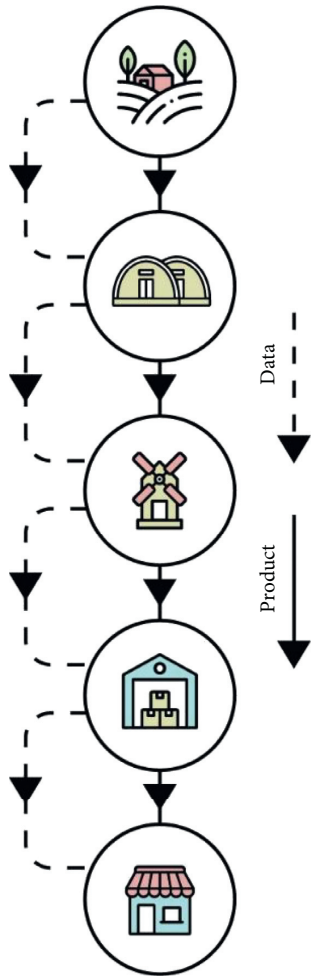

Figure 3: The traditional flow of products and associated data.

Tian analyzed the use of RFID in combination with blockchain technology and the possible advantages and disadvantages of building an agri-food chain based on the RFID and blockchain [38]. He concluded that the RFID and blockchain enhanced the food safety and efficiency of the supply chain. Moreover, a significant reduction in RFID application prices will increase the use of RFID technology in logistics over multiple times. In another article, he presents a case study of building a food supply chain using the blockchain and IoT and demonstrates the hazard analysis and critical control points identification in real-time food-tracing scenario [39].

Ramundo et al. [40] present the potential of using the state-of-the-art emerging technologies to build the food supply chain. They also evaluate the use of IoT platforms in food supply chains. They analyzed the use of IoT technologies in the farm, processing, logistics, and distribution. They concluded that many companies adopt the already existing technologies but are continually experimenting with new technologies, such as IoT, to stay at the top in the global market.

A blockchain-based traceability system for the wine supply chain is proposed by Biswas et al. [41]. The system traces the wine supply chain from grapes to bottle.

Tse et al. [42] presents the concept of blockchain technology and its potential use in information security of the food supply chain when compared to the traditional centralized supply chain.

Casino et al. briefly described the use of the blockchain into various fields of supply chain management, trading, business, and transaction settlements [43].

Baralla et al. [44] describe the use of Hyperledger Sawtooth to propose a framework to trace and secure the food supply chain based in Europe. The authors used a theoretical approach and concluded that blockchain technology is highly useful for government officials to track and audit the food supply chains.

Chen et al. [45] presented the challenges in the adoption of blockchain technology for food supply chains. Olsen et al. [46] analyzed the cost benefits and limitations of applying blockchain technology in the food industry. Johnson [47] concludes that the blockchain technology holds the strength to regulate the food industry to prevent foodborne illness. Mondal et al. [48] proposed an architecture for the food supply chain based on the Internet of Things to trace each packet of food within the supply chain in real time.

## 4. Methodology

This work is based on a design-based research approach and the concept of mindful use of information technology. The mindful use of technology focuses on using the most effective and cost-efficient features of a technology to contribute to problem-solving [49]. Design-oriented methodology focuses on the analysis of practical problems from the real world by the collaboration of researchers and practitioners to develop a solution using the existing design principles and technological innovations [50]. These solutions are then further enhanced and improved by the required research and development to deploy in the production environment as a solution for defined problem. The detailed workflow of our developed methodology based on a design-based research approach is given in Figure 4.

We need to record the circulation of products and associated data in the whole food supply chain to trace the food from farm to fork. The traditional supply chain environments are based on conventional database technologies. It does not fulfill our purpose, as there is no continuous flow of information throughout the chain in these environments.

Figure 3 is a high-level illustration of the movement of the products and associated data in traditional supply chain environments. Every supplier in the traditional supply chain moves the products and related data on its own. The availability of the data from farmers to the end consumer is scarce in these cases. The distributed databases also have no consensus mechanism, and an intruder or administrator can tamper the information.

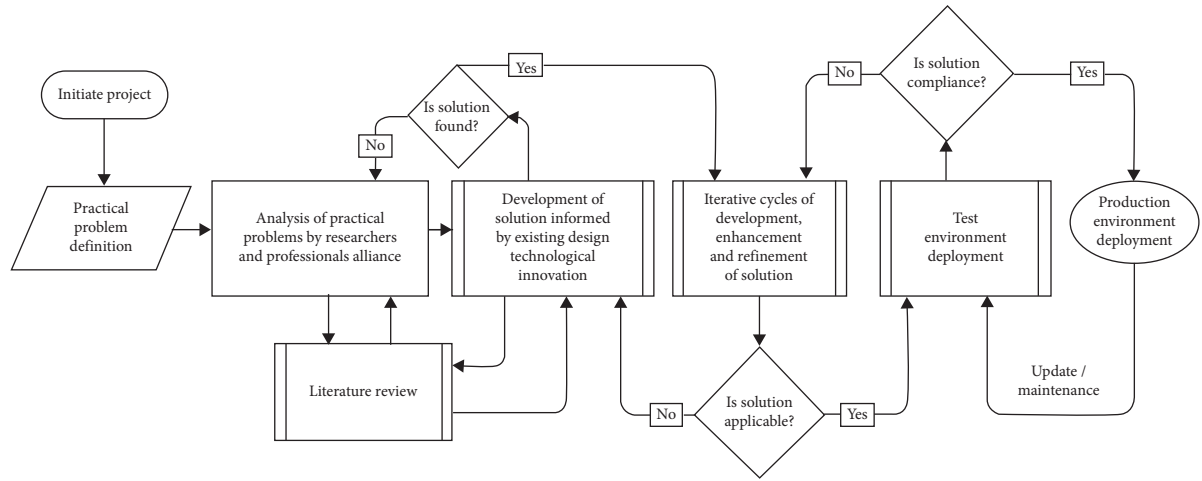

Figure 4: Research methodology.

In new emerging technologies, the blockchain is proving a potential candidate [18]. In blockchain-based supply chains, a single ledger is shared among all the entities of the system. The end buyer has the same data shared by the initial supplier plus the additional data being added at each stage of the production. Figure 5 is showing high-level details of our potential design to flow the products and associated data in blockchain-based supply chains. The blockchain makes sure the availability of the tamper-proof data from each stage of production to all the stakeholders.

To identify the product throughout the supply chain, the Internet of Things (IoT) is almost standard [51]. Still, it failed to stop the sale of horse meat as beef in the traditional supply chain environment due to the lack of continuous data flow throughout the supply chain [30]. The emerging of IoT with the blockchain is becoming a hot area of technological development.

This methodology provides a heaven for small and medium enterprises that cannot invest much to obtain or developed new technologies. This work fits best under design-based research methodology as the research is intended to solve a real-life problem by the construction of a new artifact using the already existing technology, and our consortium has members from all required fields. The next section briefly describes the blockchain and IoT technology with its core components important to this work.

## 5. Traceability System Design and Implementation

United Nations' Global Compact Office defines the three traceability models for tracing products in supply chains: the segregation, mass balance and book and claim model, drawn in Figure 6. The product segregation model makes sure the separation of certified products from noncertified products throughout the supply chain. The mass balance model allowed the mixing of certified materials with noncertified materials in a controlled manner, and certified input should not be less than certified output. The book and claim model relies on the link between the volumes of the certified material produced at the beginning of the supply chain and the number of accredited products sold at the end of the value chain [15]. This work aims to implement a product segregation model using blockchain technology and integrate it with already existed infrastructure at organizations without disturbing the traditional business practices to a large extent. We choose the Hyperledger Fabric as a blockchain platform. Hyperledger Fabric is a specialized platform for enterprises to create their own blockchain. The important features of Hyperledger Fabric are described in the coming section. Hyperledger Fabric is an open source platform and described itself as "an open source collaborative effort created to advance cross-industry blockchain technologies. It is a global collaboration, hosted by the Linux Foundation, including leaders in finance, banking, Internet of Things, supply chains, manufacturing, and technology."

*5.1. Hyperledger Fabric, an Enterprise-Ready Blockchain Solution.* The food industry is deploying information technology to capture the market share and increase the customer's trust. The distribution companies are very conscious of their potential customers. Any food safety disaster or substandard product can affect their reputation, and they can lose customers. A group of organizations includes food manufacturers, food distribution, and livestock farms association and joins a consortium to improve data sharing, production quality, supply chain, customers trust, and market share [16]. The members of a consortium are not limited to industry only. It may also contain some regulatory authorities such as the Food and Drug Administration Authority and even the animal diet producers and monitoring organizations. The typical benefits of the consortium include but not limited to standardization, collaboration, and efficiencies.

Some most essential components and services provided by Hyperledger Fabric, an open-source blockchain platform, for enterprises which made it a strong candidate of choice for this work are described below [52, 53]. Hyperledger Fabric is a programmable network that encapsulates the business logic implementation and application of the business network by way of smart contracts or chaincodes [54].

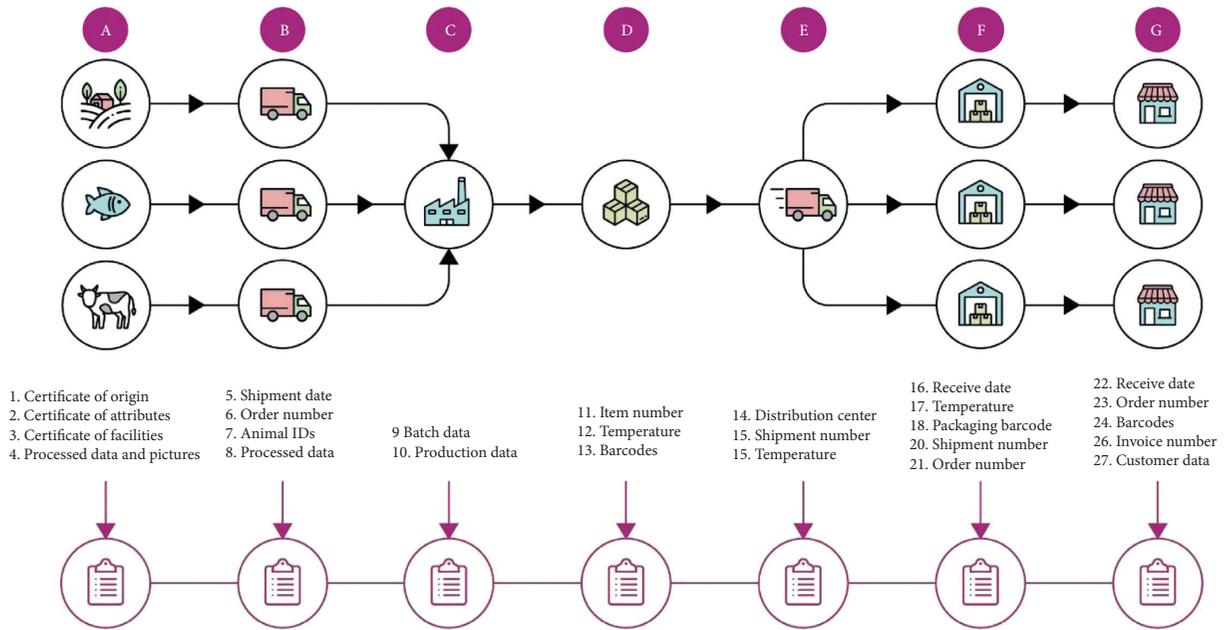

FIGURE 5: Flow of products and associated data on the blockchain.

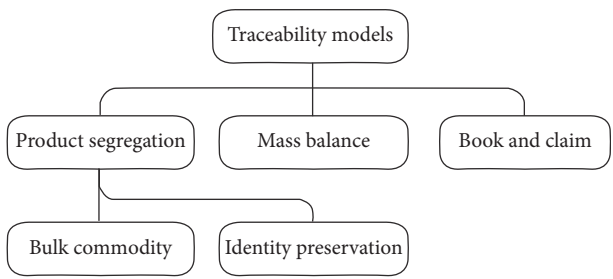

FIGURE 6: Traceability models to trace products in supply chains.

*5.1.1. Permissioned Network.* In a public blockchain network, the users downloaded the software and started to make transactions without disclosing their identities. In business networks, it is not an appropriate way of things work. The anonymity is not acceptable in business networks. The members in business networks always have a known identity and assigned roles. The Hyperledger Fabric is a permission-based network, and it assigns known identities and roles to carry out transactions. All the users and components in the Hyperledger Fabric network are needed to be identified on the network. These entities are assigned by the identities on the network by the Membership Service Providers (MSPs) and Certification Authority (CA) by the Hyperledger Fabric using the Public Key Infrastructure (PKI) to authorize and validate the users and components.

*5.1.2. Confidential Transactions.* In the business world, confidentiality from unrelated parties is a critical feature in many scenarios. Sometimes, the business networks want to keep their transactions very confidential from the unnecessary parties and only reveal them to the counterparty. Hyperledger Fabric provides the channel functionality to achieve transaction privacy between selected parties only. Each channel has one ledger, and there may be multiple channels between consortium members on the same network.

*5.1.3. Consensus and Policy Support.* Members of the consortium make many policies, decisions, rules, and regulations to run the consortium. Typically, consortium uses a decentralized approach for making decisions. Many administrators from member organizations decide with the majority to make any changes into the network, which affects the members of the consortium or business network. Such a decentralized decision system needs governance and decision-making models. Hyperledger Fabric technology supports decentralized administration by way of policies.

*5.1.4. Identity Management.* Hyperledger Fabric uses PKI (Public Key Infrastructure) for managing identities. It provides two tools: the active directory integration and Fabric-CA Server for identity management. The typical process of generating identities is given in Figure 7. The proof of identity is furnished by the identity owner to the registration authority. The registration authority validated the user information and passed it to the certification authority. The certification authority creates an x509 certificate and gives it back to the owner and validation authority to prove its validity. The other components, such as peers and orderer nodes, also require the identity to participate in the network.

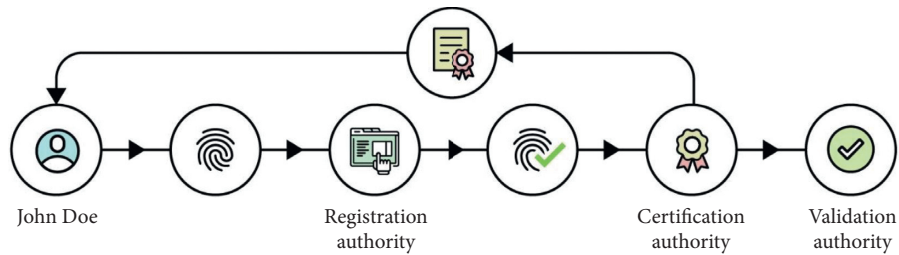

FIGURE 7: Typical process to issue x509 certificates.

*5.1.5. Application Development and Integration.* It is relatively easy to integrate the Hyperledger Fabric blockchain with the existing enterprise system. Every organization in the network can develop the interaction system according to its needs. Fabric front end applications can be developed independently using RESTful APIs as middleware, or the custom middleware can also be designed using one of the SDKs provided by Hyperledger Fabric as shown in Figure 8.

*5.2. System Architecture.* Figure 9 is showing the system architecture of this solution. This system is initially installed at three locations. The two peer nodes are installed at the livestock farms organization premises. The food-manufacturing organization and the food distribution organization each holds two peer nodes. The food-manufacturing organization additionally holds the two orderer nodes and Kafka clusters. This system is currently demonstrating Kafka as an ordering service. Hyperledger Fabric also supports Raft-based consensus. The Raft-based ordering service provides crash fault tolerance such as the Kafka ordering service implementation, without the need to manage external dependencies. Additionally, when using Raft, ordering service nodes can be provided by different organizations across the world in various data centers. The anchor peers of each organization communicate with the orderer and other peer nodes using the Internet while the Kafka brokers are connected with orderer nodes over a simple layer 2 switch.

Figure 10 is demonstrating the general application server components at each location. The application server based on NodeJS, which hosts Fabric SDK and NodeJS Express. The NodeJS Express is providing RESTful APIs and various functions to send, receive, and access data from existing IoT and software infrastructure at the organizations. The IT support teams from all the organizations engaged during this development and deployment. This application server provides a generic interface that can use to generate a new user interface or integrate it into an already existing application interface. Every current organization or future organization joining this venture can integrate this solution into their existing system according to their needs with very minimum efforts due to its simplicity.

*5.3. Identity Management at Livestock.* Our farms are already certified as organic product producers. The animals at the farm are marked and injected with RFID tags. All the related parameters of the animals are structured. This tag contains the necessary information stored in it, such as the date of birth, breed, and ownership information. The unique identification (ID) of the tag is the identity of the animal, or the manual ID can be assigned. The veterinarian scans and puts all the information regarding disease control, vaccination, and weight in the system. It is also tracking and recording the conditions of the production site, which includes the environmental conditions including water quality, water temperature, air quality, environment temperature and humidity, labor conditions, and the quality of the processes. The data from the animal tags and environmental sensors including veterinary services are collected by installed antennas in the farms and send to the workstation at the facility as well as to the enterprise HQ owning the farms as demonstrated in Figure 11. The system stores those data in the traditional databases for internal use alongside posting these transactions on the blockchain. These certified organic farms work on two models. They raised their batch of animals or only provided services to food-manufacturing organizations. In both scenarios, the related data are captured and stored according to the organic certification authority. These data are transferred to the customers with the product as product specifications. The blockchain application is integrated in such a way that it does not affect any existing system.

The system uses a chaincode to store data on the blockchain. It only transfers the required data at a fixed frequency from the current traditional database to the blockchain using that chaincode. A new version of chaincode deployed for each new batch of animals. The chaincode is required to put or read the data from the blockchain. These chaincodes are transferred to the customer with the product as product specifications.

The farms also deliver organic meat instead of live animals. In this case, a separate chaincode is used to manage the unique identity of each meat package. The farm generates the set of new identities for each package of meat against the animal identity as illustrated in Figure 12. These identities are stored on the blockchain from the system using a separate chaincode developed explicitly for this purpose. This new chaincode reads the animal's identities against each transaction using the previous-stage chaincode and stores this set of identities for this product. Both chaincodes transfer with the product to the customer as product specifications. It makes each meat package traceable with the complete history of the animal.

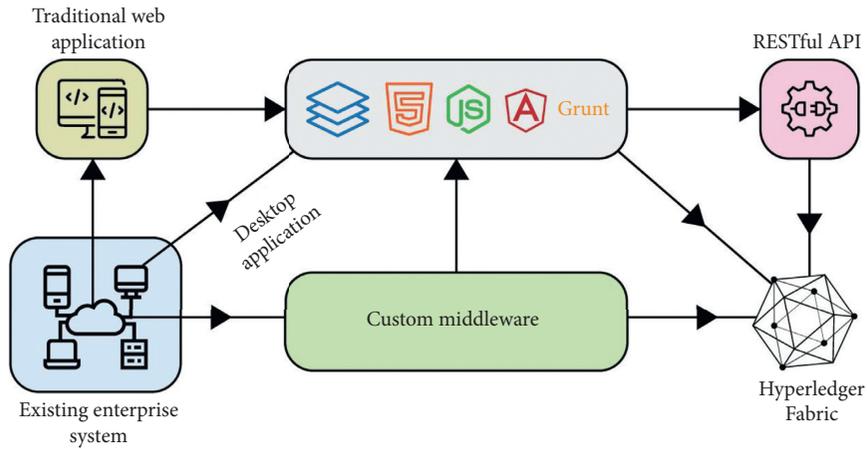

FIGURE 8: Hyperledger Fabric application development.

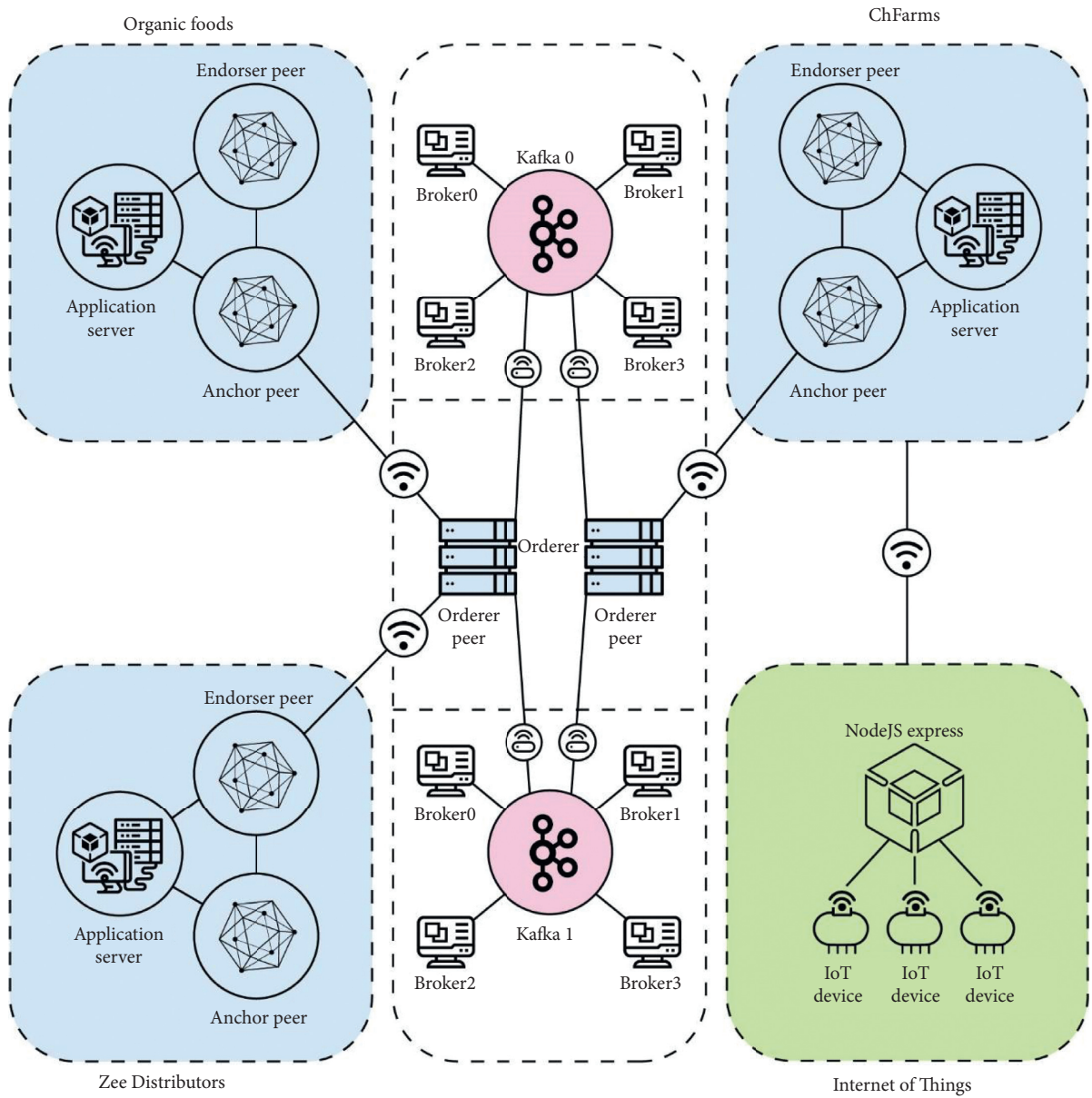

FIGURE 9: An overview of the system.

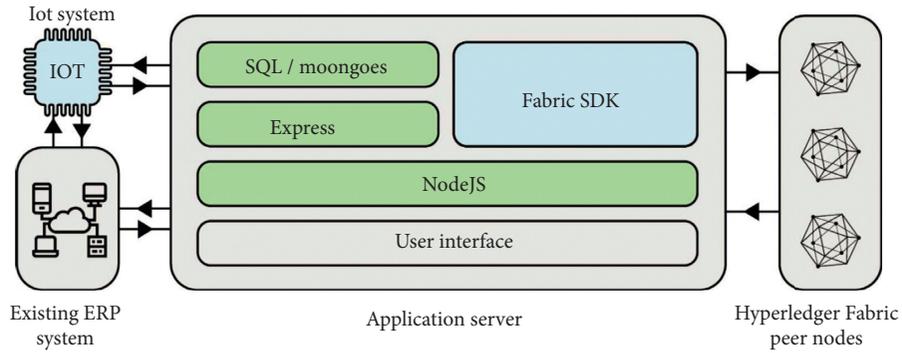

FIGURE 10: Application server.

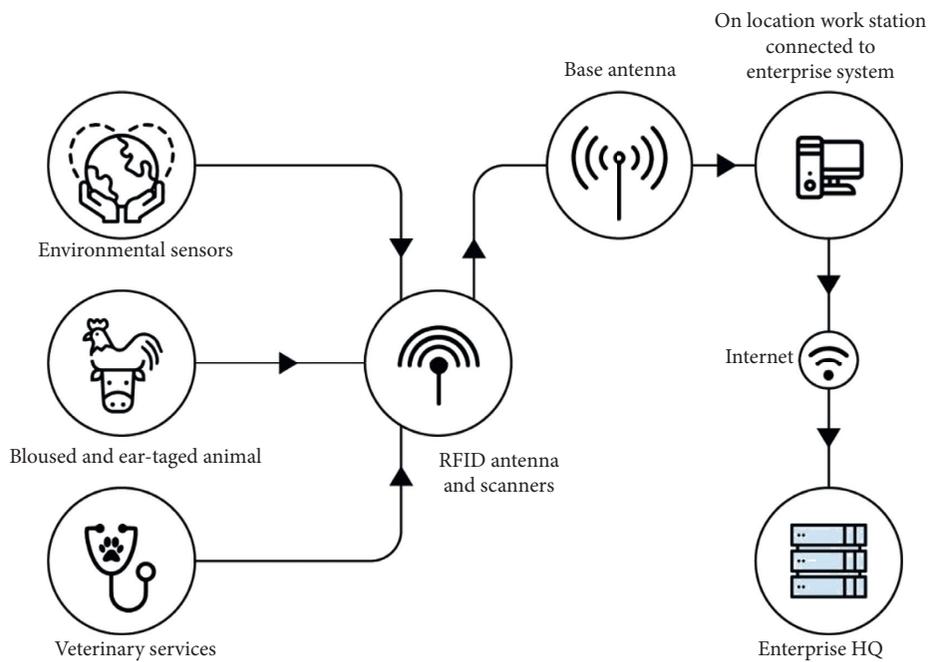

FIGURE 11: IoT-enabled farming.

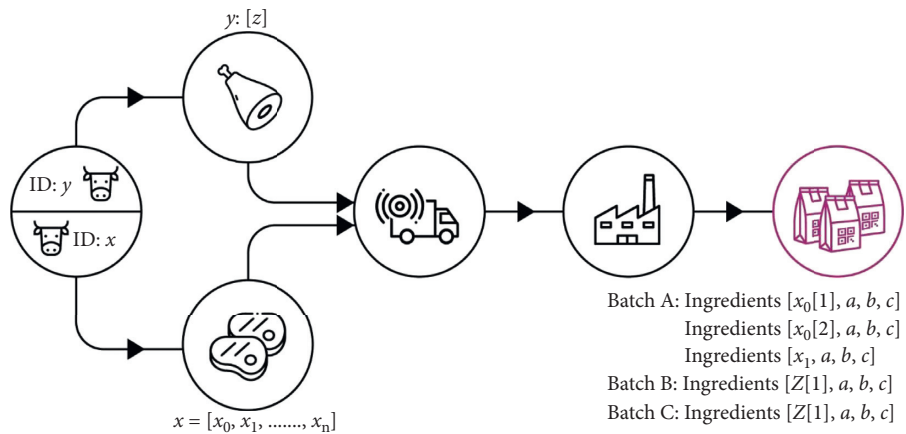

FIGURE 12: Product identification management throughout the supply chain.

*5.4. Identity Management at Food-Manufacturing Plant.* The food package can be traced and tracked into the supply chain from farms to a food processing company to the distributor, distributor to the whole seller, and whole seller to the retailer. It is possible with the help of bar code, QR code, or RFID tag on each box of food. The blockchain can easily handle the identities and regulate who gets access to the information behind each product. Each box of meat contains an RFID or QR code which provides information about it, and it is possible to track that animal and its complete production process from birth to meat processing as shown in Figure 12. The shipping company is providing real-time data from its shipping trucks using its IoT platform, which is accessible by both the shipper and receiver. These data are also updating through a smart contract at specific intervals on the blockchain.

*5.5. Food-Manufacturing Plant.* The food manufacturer receives its raw material shipment with its product specification chaincode. This chaincode enables the manufacturer to read the raw material specifications from the blockchain. The manufacturer transfers these data from the blockchain to its traditional database management system for its internal operations and records. If the company receives a live animal, it deploys the smart contract to update the set of identities against each package of food the same as a livestock farm.

The organization deploys a chaincode to record its production on the blockchain. This chaincode queries the raw material using the raw material specification chaincode it receives with its raw material. The organization system assigns a unique identification number to finished product against the ingredients identification group. Each piece of meat package has its ID attached in slaughterhouse against animal ID. Each box of meat is scanned, and the information is entered into the database. A set of new sub IDs created by the system against each ID, assigned to each batch of products where the meat is in use, as shown in Figure 13. That identification also prints on the finished product in the form of QR or barcode. These identities are transferred from traditional databases to the blockchain using an explicitly developed chaincode for a batch of production. These chaincodes later deliver as a final product specification to the distributor with the product delivery. A mobile phone application is also under development for consumers where they can trace the whole supply chain by providing the identification of the finished product.

## 6. Hardware Cost

This solution is cost efficient for small and medium organizations that cannot invest in building a whole new system from scratch. The Hyperledger Fabric is an open-source platform that receives regular updates. We used existing organizations' resources. All the nodes deployed on the 4th generation intel core i7 processor consists of 32 GB of RAM and 2 TB of storage media.

The system does not interfere with existing infrastructure but provides a separate layer of data sharing that can be integrated into the system using a generic application structure as a reference. In each stage of production, the organizations are using their traditional database systems or transfer the data from the blockchain to the system for their business tasks. The blockchain is used only as a tool to make a record of transactions that cannot be altered and transferred to the customer.

This blockchain-based data sharing system is expandable as any organization can join or leave the blockchain consortium at any time. The new coming organizations do not need to change or develop the system according to the consortium need. They can integrate it without much extra development cost.

## 7. Challenges and Obstacles

There are many blockchain applications for the food supply chain under trial or implementation worldwide. Companies and organizations are facing several challenges and obstacles in blockchain adoption. Two major challenges and obstacles are described in this section.

*7.1. Interoperability and Universal Acceptance.* There is not any formal road map to adopt the blockchain as global tool for transactions. The blockchain-based traceability for the food supply chain is not even yet ready to implement as a nationwide uniform system. In many countries, the rules, regulations, standards, and labeling are conflicted between the states. There is no technology exist to connect different blockchain systems. This lack of interoperability and data sharing is currently the biggest obstacle. The blockchain system compared to the traditional electronic traceability system is more homogeneous as it stores the transactions instead of state values which makes interoperability and sharing of data easier than the traditional traceability systems. The states and countries need to develop a standard ontology which defines what the recorded data elements and values mean in the food supply chain. They also need common messaging standards.

*7.2. Laws and Regulations.* The attempts to regulate the blockchain have been another area of controversy. The role of regulators and state authorities are not clear. The functioning of blockchain may also conflict with regulatory requirements. For instance, information shared in a ledger cannot be modified or altered. This feature is in contradiction with the right to forgetting in many countries.

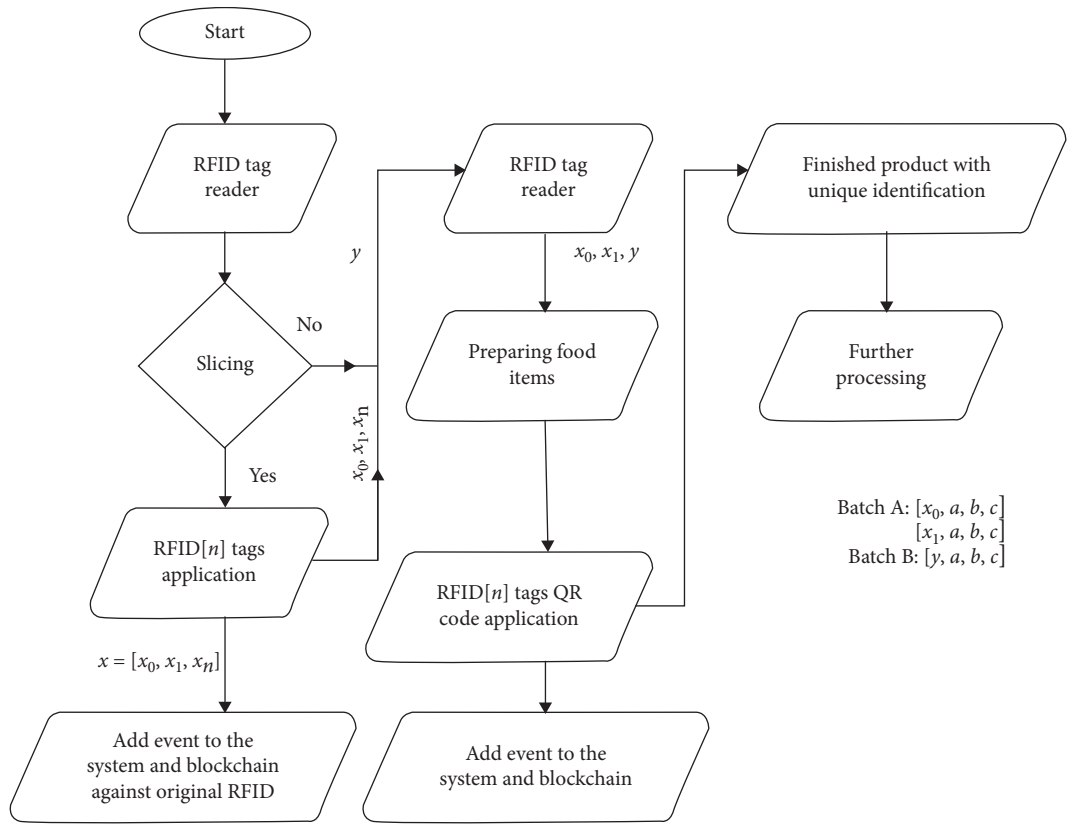

FIGURE 13: Flowchart explaining the identity generation process.

## 8. Conclusion

This paper applies blockchain technology in the food industry in a holistic and mindful approach to measuring and certifying the quality of the final product cost-effectively. It helps to provide transparent information about the food from farm to fork. This information is also useful for Food Control Authorities to prevent potential food safety hazards. It also enhances healthy competition between companies to improve product quality continuously. It provides decentralized, credible, and transparent information, being stored on the blockchain parallel to the traditional database system behind the scene with existing enterprise systems without any noticeable change in conventional business operations and training general employees.

## 9. Future Work

We are working with food certification authorities and NGOs to unify the type of data (ontology) that are required to certify and record for organic products within China. It will help to create a unified chaincode for different stages of the supply chain. This will provide a cost-efficient and robust system with a unifying standard API.

Many governments are also showing interest in using blockchain technology in governance. It needs to develop a system where the different blockchains can transfer the data on each other or can be merged. No business can be fully centralized or decentralized without compromising in areas such as security, privacy, performance, and scalability. Governments and private organizations are also concerned about their data privacy and protection. It needs to develop a mechanism to process the data on their servers at their premises and bring only desired results from the servers to the blockchain for analysis and predictions. The biggest hurdle for governments to be part of this chain is the laws to protect the data that can be used to affect fair competition and other national food security threats.

## Data Availability

Data sharing is not applicable for this article.

## Conflicts of Interest

The authors have no affiliation with any organization with a direct or indirect financial interest in the subject matter discussed in the manuscript.

## Acknowledgments

The authors would like to acknowledge the support provided by the National Key R&D Program of China (no. 2018YFC1604000).